\documentclass[trackchanges]{aastex63}

\usepackage{CJK}
\usepackage{amsmath}
\usepackage{gensymb}
\usepackage{amssymb}
\usepackage{hyperref}
\usepackage{tikz}
\usetikzlibrary{arrows}
\usepackage{booktabs}
\usepackage{multirow}
\usepackage{tabularx}
\usepackage{xcolor}
\usepackage{booktabs,multirow}

\definecolor{DarkGreen}{rgb}{0.0,0.45,0.0}  % define a custom color
\newcommand{\B}[1]{{\color{blue}{#1}}}      
\newcommand{\R}[1]{{\color{red}{#1}}}     
\newcommand{\G}[1]{{\color{DarkGreen}{#1}}}
\newcommand{\M}[1]{{\color{magenta}{#1}}}
\shorttitle{Filament Chirality, Sense of Rotation, and Hemispheric Preference}
\shortauthors{Zhou et al.}

\begin{document}

\begin{CJK*}{UTF8}{gbsn}

\title{The Relationship between Chirality, Sense of Rotation, and Hemispheric Preference of Solar Eruptive Filaments} 
\correspondingauthor{Zhenjun Zhou, Rui Liu}
\email{zhouzhj7@mail.sysu.edu.cn, rliu@ustc.edu.cn}

\author[0000-0001-7276-3208]{Zhenjun Zhou(周振军)}
\affiliation{School of Atmospheric Sciences, Sun Yat-sen University, and Key Laboratory of Tropical Atmosphere-Ocean System, Ministry of Education, and Southern Marine Science and Engineering Guangdong Laboratory (Zhuhai), Zhuhai, China}
\affiliation{CAS Key Laboratory of Geospace Environment, Department of Geophysics and Planetary Sciences, University of Science and Technology of China, Hefei, Anhui 230026, China}
\affiliation{CAS Center for Excellence in Comparative Planetology, China}
\author[0000-0003-4618-4979]{Rui Liu}
\affiliation{CAS Key Laboratory of Geospace Environment, Department of Geophysics and Planetary Sciences, University of Science and Technology of China, Hefei, Anhui 230026, China}
\affiliation{CAS Center for Excellence in Comparative Planetology, China}
\author[0000-0003-2837-7136]{Xing Cheng}
\affiliation{School of Astronomy and Space Science, Nanjing University, Nanjing 210023, China}
\author[0000-0002-7018-6862]{Chaowei Jiang}
\affiliation{Institute of Space Science and Applied Technology, Harbin Institute of Technology, Shenzhen 518055, People's Republic of China}
\author[0000-0002-8887-3919]{Yuming Wang}
\affiliation{CAS Key Laboratory of Geospace Environment, Department of Geophysics and Planetary Sciences, University of Science and Technology of China, Hefei, Anhui 230026, China}
\affiliation{CAS Center for Excellence in Comparative Planetology, China}
\author[0000-0001-6804-848X]{Lijuan Liu}
\affiliation{School of Atmospheric Sciences, Sun Yat-sen University, Zhuhai, Guangdong, 519000, China}
\affiliation{CAS Center for Excellence in Comparative Planetology, China}
\author[0000-0002-4721-8184]{Jun Cui}
\affiliation{School of Atmospheric Sciences, Sun Yat-sen University, Zhuhai, Guangdong, 519000, China}
\affiliation{CAS Center for Excellence in Comparative Planetology, China}

\begin{abstract}

The orientation, chirality, and dynamics of solar eruptive filaments is a key to understanding the magnetic field of coronal mass ejections (CMEs) and therefore to predicting the geoeffectiveness of CMEs arriving at Earth. However, confusion and contention remain over the relationship between the filament chirality, magnetic helicity, and sense of rotation during eruption. To resolve the ambiguity in observations, in this paper, we used stereoscopic observations to determine the rotation direction of filament apex and the method proposed by \cite{Chen_etal_2014} to determine the filament chirality. Our sample of 12 eruptive active-region filaments establishes a strong one-to-one relationship, i.e.,
during the eruption, sinistral/dextral filaments (located in the southern/northern  hemisphere) rotate clockwise/counterclockwise when viewed from above, and corroborates a weak hemispheric preference, i.e., a filament and related sigmoid both exhibit a forward (reverse) S shape in the southern (northern) hemisphere, which suggests that the sigmoidal filament is associated with a low-lying magnetic flux rope with its axis dipped in the middle. As a result of rotation, the projected S shape of a filament is anticipated to be reversed during eruption.  
\end{abstract}

\section{Introduction}\label{sec:intro}
Solar filament eruptions, flares, and coronal mass ejections (CMEs) are closely associated with each other, usually originating from the same polarity inversion line (PIL) of an active region \citep{Green2018}. Aligned along the PIL, a cold H$\alpha$ filament and/or a hot sinuous loop termed sigmoid are often identified in the corona \citep[e.g.,][]{Pevtsov_2002}. These structures are thought to characterize a current-carrying (sheared or twisted) magnetic field system possessing significant magnetic helicity \citep{Pevtsov2014}. CMEs are the key drivers of adverse space weather events, mostly because they can provide sustained periods of strongly southward magnetic field, allowing efficient transportation of solar wind energy, plasma, and momentum into the Earth's magnetosphere \citep{Kilpua2019}. Hence, the geoeffectiveness of a CME arriving at Earth depends very much on the orientation and helicity of its magnetic structure \cite[e.g.,][]{Yurchyshyn2001,Awasthi2018} . 

To understand the magnetic field structure and orientation of a CME, one  has to resort to the photospheric magnetic field of its source region as well as its initial formation and evolution in the low corona, due to the absence of magnetic field measurements in the corona. In this regard, filaments and sigmoids are two most important CME progenitors \citep{Gilbert2000,Canfield1999,Liu2010}. However, how a progenitor evolves into a CME and eventually into an interplanetary ejecta remains elusive, which is further complicated by the rapid and complex rotations during the early evolution of CMEs \citep[e.g.,][]{Green_etal_2007,Vourlidas2011,Thompson2012}, although some interplanetary CMEs may still have similar axial field directions as the erupted filaments \citep[e.g.,][]{Bothmer&Schwenn1994,Yurchyshyn2001}.A few case studies and numerical simulations \citep[e.g.,][]{Green_etal_2007,Lynch2009,Torok_etal_2010} indicated that the direction of rotation depends on the chirality of CMEs' core structure---a magnetic flux rope---a right-handed (left-handed) flux rope rotates clockwise (counterclockwise). From the observational perspective, however, the chirality of coronal structures is often ambiguous, because what one typically sees in an image of the corona is all the coronal emission along the line of sight collapsed and projected to the plane of sky. For example, \citet{Green_etal_2007} employed five different ``helicity indicators''; although in their study of four events these indicators generally agree with each other, the individual method seems insufficient to determine the helicity sign of eruptive structures. 

The chirality of filaments and sigmoids are ambiguous in their own ways. 
The majority of sigmoids are found to be forward (reverse) S-shaped in the southern (northern) hemisphere, following a hemispheric helicity rule that is independent of solar cycle, i.e., positive (negative) helicity is preferred in the southern (northern) hemisphere \citep{Pevtsov2014}. However, the empirical rule which associates forward (reverse) S shape with positive (negative) helicity may be only applicable to low-lying structure, because a high-lying loop could have opposite sign of writhe as a low-lying one, even though both are projected into the same S shape on the disk \citep{Torok_etal_2010}. 

Similarly, it is believed that there is a one-to-one correspondence between filament chirality and magnetic helicity sign \citep{Rust_1999}. By definition, a filament is dextral (sinistral) if its axial magnetic field points right (left) when viewed from its positive-polarity side of the PIL \citep{Martin1998}. \citet{Martin_etal_1992} suggested that a dextral (sinistral) filament has right-bearing (left-bearing) barbs, a bundle of filament threads extruding out of the filament spine in a way similar to right- or left-bearing exit ramps off a highway. However, the bearing sense of filament barbs could be ambiguous because of foreshortening and projection effects; moreover, barbs developed in a magnetic sheared arcade are expected to be oriented oppositely to those in a magnetic flux rope \citep{Guo_etal_2010,Chen_etal_2014}. Alternatively, \citet{Chen_etal_2014} proposed to determine the filament chirality based on the skewness of the conjugate filament drainage sites with respect to the PIL during the filament eruption. Unambiguously, an eruptive filament with right-skewed (left-skewed) drainage sites is sinistral (dextral), corresponding to positive (negative) helicity. Using this method, \citet{Ouyang_etal_2017} found 91.6\% of 571 eruptive filaments follow the hemispheric helicity rule in comparison to 66\%-86\% using Martin's rule \citep[e.g.,][]{Pevtsov_etal_2003}. 

Filaments can also be S-shaped. Such a sigmoidal filament is often observed underneath a co-spatial sigmoid and to survive the latter's eruption \citep{Gibson2002,Pevtsov_2002,Cheng_etal_2014}, indicating the presence of a double-decker structure \citep{Liu_etal_2012,Cheng_etal_2014}. It is also noticed that an erupting filament which is S-shaped before the eruption could reverse the sense of its S shape through rotation during eruption \citep{Rust&Labonte2005,Romano2005,Green_etal_2007}. This is interpreted as a signature of the conservation of helicity when a writhed flux rope transforms from having a dipped central section to a humped one while maintaining the sign of writhe \citep{Torok_etal_2010}. Obviously, such complex 3D structures as well as their sense of rotations can not be validated without stereoscopic observations \citep{Zhou_etal_2017}.

The Solar Terrestrial Relations Observatory \cite[STEREO;][]{Kaiser2008}, aided by ground-based telescopes and space-born instruments operating at the Earth orbit such as the Solar Dynamics Observatory \citep[SDO;][]{Pesnell_etal_2012} and Hinode \citep{Kosugi2007}, has made it possible to observe the Sun with high-resolution high-cadence images taken from as many as three viewing angles. This gives us an unprecedented advantage to resolve the above-mentioned ambiguity as compared with previous studies. Here we revisit the problem of filament rotation in relation to chirality by applying the method proposed by \citet{Chen_etal_2014} to determine the filament chirality and by investigating the filament shape and rotation direction with multi-viewing-angle observations. In the rest of the paper, we present our analysis of the observations in \S\ref{sec:Observation} and make concluding remarks in \S\ref{sec:Results}.

\section{Observation \& Analysis}\label{sec:Observation}

\subsection{Instruments}\label{subsec:Instruments}

In this study we used H\(\alpha \) images from the Global Oscillation Network Group \citep[GONG;][]{Harvey_etal_2011} and cool EUV passbands onboard SDO and STEREO, such as 304 {\AA} (\ion{He}{2}), to study the filament morphology and evolution. Filaments are optically thick in 304 {\AA} and H$\alpha$, hence largely free from the line-of-sight confusion. But during eruption, they become less dense due to expansion, which combined with the projection of complex dynamic motions often introduces confusion for a single viewing angle. Hence, stereoscopic observations are indispensable in determining the sense of filament rotation. 

The associated sigmoids are best visible in soft X-ray images taken by the X-Ray Telescope \citep[XRT;][]{Golub_etal_2007} onboard Hinode. When XRT data are not available, we resorted to the Atmospheric Imaging Assembly \citep[AIA;][]{Lemen_etal_2012} onboard SDO. Among AIA's EUV passbands 335~{\AA} is similar to XRT with its temperature response function peaking at about 2.5 MK \citep{Savcheva_etal_2014}. 

To get the context of magnetic environment, we used vector magnetograms provided by the Heliospheric and Magnetic Imager \citep[HMI;][]{Schou_etal_2012} onboard SDO. The $180\degree$ ambiguity of the transverse field is resolved by an improved
version of the minimum energy method \citep{Leka_etal_2009}.

\subsection{Selection of Events} \label{subsec:Selection}
We examined filament eruptions recorded by both SDO and STEREO from May 2010 to Sep 2019, and selected 12 filament eruption as listed in Table~\ref{tab:mytable} according to the following criteria: (1) the rotation direction of the eruptive filament in the low corona can be unambiguously determined by SDO and STEREO observations; and (2) the filament chirality can be unambiguously determined by the skewness of drainage sites relative to the PIL \citep{Chen_etal_2014}. Details of the selected events are available at this \href{http://sysu-pearl.cn:8080}{catalog website}\footnote{\url{http://sysu-pearl.cn:8080}}.

\subsection{Morphology, Sense of Rotation, \& Filament Chirality } \label{subsec:Method}

First we determined whether the filaments and sigmoids present forward (Column 4) or reverse S shape (Column 5 of Table~\ref{tab:mytable}). Keeping in mind that filament material does not necessarily occupy the whole length of magnetic flux tubes, we looked through H\(\alpha \) and 304~{\AA} images within one day prior to eruption to make out the complete shape of the filament. Take Case \#5 in Table~\ref{tab:mytable} as an example. The dark feature observed by GONG H\(\alpha \) telescope outlines a reverse S-shaped filament (Fig.~\ref{fig:figure1}a), which looks more prominent in AIA 304 {\AA} (Fig.~\ref{fig:figure1}c). In spite of relatively low resolution for this full-disk image, Hinode/XRT also shows a reverse S-shaped sigmoid. 

Next we determined the rotation direction of the filament apex during eruptions (Column 6 of Table~\ref{tab:mytable}). In the same example as above, SDO and STEREO-A are separated by 114$\degree$. From the top view by SDO (Fig.~\ref{fig:figure1}c and accompanying animation), one can see that the apex of the filament initially displays counterclockwise rotation, but it soon becomes too rarefied to be seen. On the other hand, from STEREO-A's perspective, one can see that originally the filament takes a semi-circular shape above the limb (00:36 UT; Fig.~\ref{fig:figure1}d), and then transforms to an inverse $\gamma$-shaped configuration (00:46 UT; Fig.~\ref{fig:figure1}d). There may be confusion as to whether its southern leg is in front of or behind the northern one in this inverse $\gamma$ configuration, due to the low cadence, but combine the two viewing angles, one can determine without ambiguity that the filament rotates counterclockwise while rising. The rotation direction of the rest events is similarly determined. Note there is no STEREO data for Case \#12, but SDO provides a clear view of the clockwise rotation, with the filament located near the disk center and erupting radially, similar to Case \#5 (Fig.~\ref{fig:figure1}).

To determine the filament chirality (Column 3 of Table~\ref{tab:mytable}), we relied on filament material falling down along the two legs of a filament during eruption to locate its magnetic footpoints. The impact at the surface often yields EUV brightening due to compressional heating of the falling material. However, the eruptive filament may also interact with ambient coronal magnetic structures to produce remote impact sites or even remote flare ribbons in UV emission; these are caused not only by the redirection of falling material towards remote magnetic footpoints, but by the field-aligned transport of energy released by magnetic reconnections \citep[see][and references therein]{Liu2018}. Here we focused on a pair of major draining sites near the filament channel, which are supposed to be the conjugated footpoints of the pre-eruptive filament. Thus, a filament is dextral (sinistral) when the conjugated draining sites are left-skewed (right-skewed) with respect to the PIL \citep{Chen_etal_2014}, as illustrated in Fig.~\ref{fig:figure2}(a \& b). Usually the draining sites are co-spatial with two ends of a sigmoidal filament. For example,  mapping the complete filament structure of Case \#5 (outlined by cross symbols) on an associated HMI magnetogram (Fig.~\ref{fig:figure2}c), one can see that the chirality is clearly dextral.  When the source region is behind the solar disk (Case \#3), we investigated the filament chirality several days prior to the eruption when this filament could be seen on the solar disk in the view angle of SDO. Whether these filament eruptions are associated with a corresponding CME are checked in the coronagraph observations and listed in the Column 8 of Table~\ref{tab:mytable}.

   \begin{deluxetable}{cccccccc}
   \tablenum{1}
   \tablecaption{Characteristics of 12 selected active-region filaments. Listed from left to right are the No. of events, start time of filament eruption, filament chirality, forward (reverse) S shape of the filament and the related sigmoid as indicated by `S' (`Z'), respectively, sense of rotation during eruption, hemispherical location of the source region,  and whether this filament eruption is associated with a CME (Erupted (E) or Confined (C)).}
   %Full (F), Partial (P), or Confined (C) for the success of the filament eruption.
   \label{tab:mytable}
   \tablewidth{0pt}
   \tablehead{
   \colhead{Number} & \colhead{Start time (UT)} & \colhead{chirality}&
  \colhead{Filament shape} & \colhead{Sigmoid shape} & \colhead{Filament Rotation} & \colhead{Hemisphere} & \colhead{Eruption} \\
   }
   \decimals
   \startdata
   1  & 2010-08-07 17:45    & dextral   & Z  & Z      & CCW      & N  & F  \\
   2  & 2011-06-06 04:53    & dextral   & Z  & *      & CCW      & N  & F  \\
   3  & 2011-12-25 07:56    & dextral   & Z  & *      & CCW      & N  & F  \\
   4  & 2012-05-05 17:00    & dextral   & Z  & Z      & CCW      & N  & C  \\
   5  & 2012-05-10 00:00    & dextral   & Z  & Z      & CCW      & N  & F  \\
   6  & 2012-05-22 01:39    & dextral   & Z  & Z      & CCW      & N  & F  \\
   7  & 2012-08-11 16:27    & sinistral & *  & *      & CW       & S  & C  \\
   8  & 2012-10-24 01:06    & dextral   & S  & *      & CCW      & N  & F  \\
   9  & 2012-10-25 03:16    & dextral   & S  & S      & CCW      & N  & C  \\
   10 & 2012-11-10 04:30    & sinistral & S  & S      & CW       & S  & F  \\
   11 & 2012-11-29 11:31    & dextral   & Z  & *      & CCW      & N  & C  \\
   12 & 2018-03-06 13:44    & sinistral & S  & *      & CW       & S  & F  \\ 
 \midrule  
   \enddata
   \tablecomments{`*' indicates that the sense of S shape is ambiguous.}
   \end{deluxetable}

\section{Results \& Discussion} \label{sec:Results}
A one-to-one relationship, as illustrated in Figure~\ref{fig:figure4}, stands out in our sample of 12 filament eruptions: the apex of a sinistral (dextral) filament rotates clockwise (counterclockwise) during the eruption. The chirality of the filaments as determined by the skewness of the drainage sites is fully consistent with the hemispheric helicity rule, i.e., those in the northern (southern) hemisphere are dextral (sinistral), therefore being associated with negative (positive) helicity. 
	
\begin{figure}[ht!]
\plotone{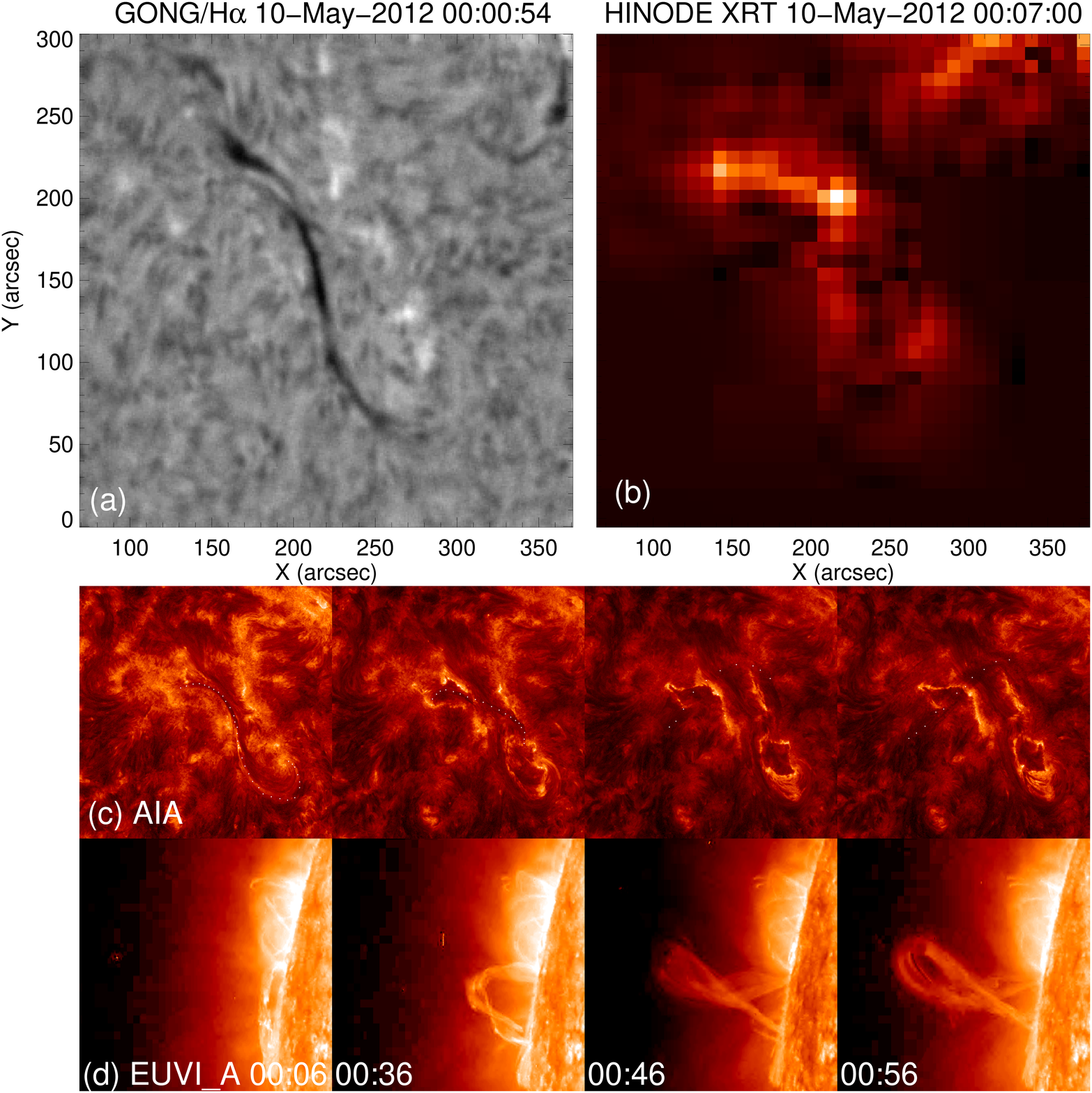}
\caption{Observation of Case \#5 listed in Table~\ref{tab:mytable}. Panel (a) shows the filament in an reverse S shape observed by GONG/H\(\alpha \) on 2012 May 10 at 00:00 UT. The corresponding sigmoid observed by Hinode/XRT is shown in (b). Panels (c) and (d) provide sequential snapshots of the filament eruption from the limb view in STEREO-A/EUVI 304 {\AA} and disk view in SDO/AIA 304 {\AA}, respectively, showing the CCW rotation of the filament. An animation of 304 {\AA} images from both viewing angles are available online. 
\label{fig:figure1}}
\end{figure}

\begin{figure}[ht!]
\plotone{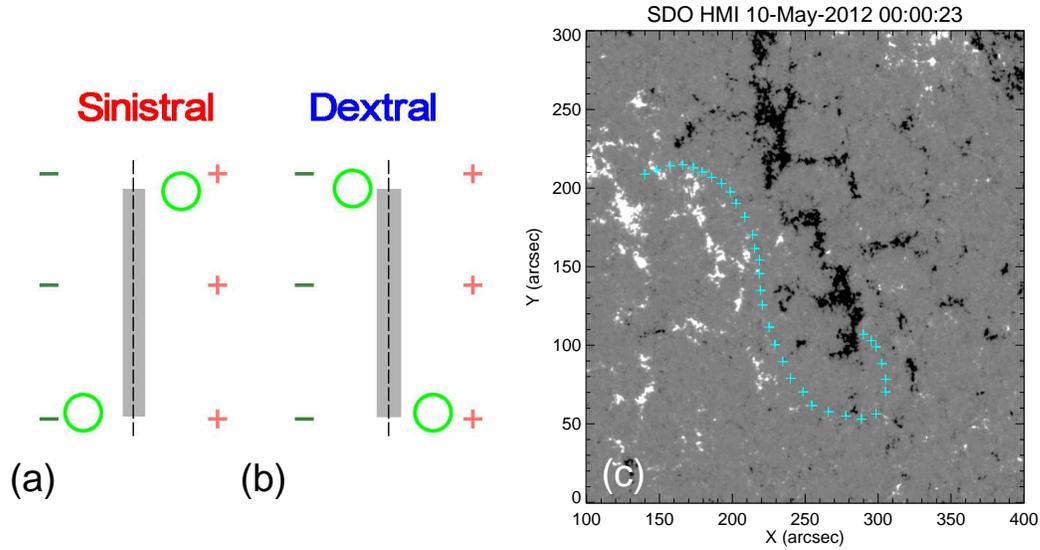}
\caption{Determination of filament chirality by the filament draining sites. Sinistral and dextral chirality are illustrated in Panels (a) and (b), respectively, with the PIL indicated by dashed line, draining sites by green circles, and magnetic polarities by `+' and `-'. In Panel (c), the cyan plus symbols outlining the EUV filament spine (Case \#5) are projected onto an HMI magnetogram of local $B_r$.
\label{fig:figure2}}
\end{figure}

\begin{figure}[ht!]
\plotone{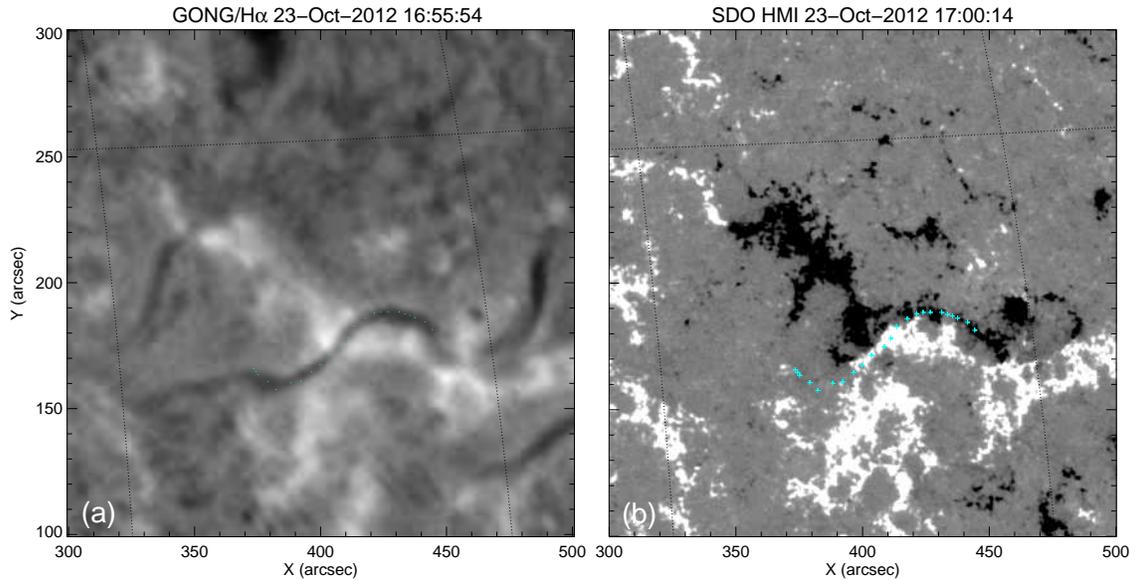}
\caption{A filament apparently violating the hemispheric rule (Case \#8). (a) displays the filament in H\(\alpha \); (b) shows the corresponding HMI magnetogram of local $B_r$.
\label{fig:figure3}} 
\end{figure}

\begin{figure}[ht!]
\plotone{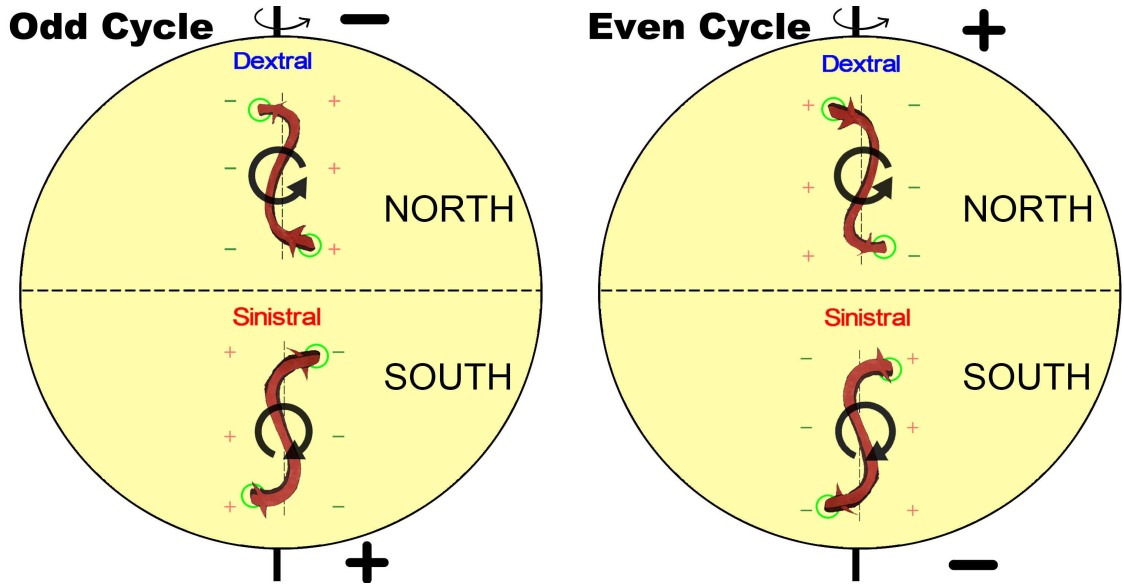}
\caption{Illustration of the relationship between the filament chirality, sense of rotation during eruption, and hemispheric preference during odd (left) and even (right) solar cycles. 
\label{fig:figure4}}
\end{figure}

In contrast, the morphological chirality, i.e, the forward or reverse S shape, is sometimes ambiguous due to the shape of the PIL. For example, in Case~\#8, the filament is dextral and located in the northern hemisphere, but bears a forward S shape. It turns out that the middle section of the PIL is distorted in a way that gives an appearance of forward S (Fig.~\ref{fig:figure3}b); in this case, the curved filament legs may be a more reliable sign of chirality \citep[see also][]{Rust_Martin_1994}. A similar situation is found in Case~\#9. The ambiguity may also result from projection as demonstrated by \citet{Torok_etal_2010}. In this regard, a sigmoidal filament is expected to reverse its S-shape in the course of the eruption to maintain the sign of its axis writhe (e.g., Figure~\ref{fig:figure1}). The reversal of the S shape may have important implications, because the original northward axial field of an eruptive structure can be rotated to be southward if the S shape is to be maintained during the interplanetary transportation. 

Based on four cases \citet{Green_etal_2007} reached a similar conclusion  that for positive (negative) helicity the apex of a filament rotates clockwise (counterclockwise). But the filament is believed to be associated with a flux rope taking on an S shape that is opposite to the corresponding sigmoid, the latter of which represents a current layer wrapping around the flux rope \citep{Titov_Demoulin_1999,Kliem_etal_2004}. \citet{Green_etal_2007} hence rejected those models implying identical orientation for the sigmoid and the flux rope \citep[e.g.,][]{Rust_Kumar_1996,Kusano2005}. But in our sample and in the literature \citep[e.g.,][]{Pevtsov_2002,Cheng_etal_2014,Zhou_etal_2017} often the cold filament and the hot sigmoid are co-spatial and have the similar S shape, suggesting that the filament is associated with a low-lying magnetic flux rope with its axis dipped in the middle \citep[e.g.,][]{Zhou_etal_2017}.

Not all these filament eruptions that are initiated make it to a full eruption, some of these cases expelled material flows back down to the Sun. Specifically,
cases \#4,\#7,\#9, and \#11 expell material flows back down the legs of an erupting prominence and have no corresponding CME in the coronagraph observations, this subset of events are known as failed eruptions or confined eruptions,
likely because of interactions with external magnetic field structures that may arrest the eruption or facilitate additional draining \citep{McCauley_etal_2015,Zhou_etal_2019}. This interaction may induce alternative rotation mechanism concerning the mutual orientation of the overlying arcade field and the erupting filament body \citep{Isenberg_Forbes_2007}. For a given chirality of the erupting field, this mechanism yields the same rotation direction as the kink instability \citep[e.g.,][]{Kliem_etal_2012}.

To summarize, by applying the method proposed by \citet{Chen_etal_2014} to determine filament chirality  and using multi-viewpoint observations to investigate the filament rotation, we established a strong relationship that during eruption sinistral/dextral filaments, which are located in the southern/northern hemisphere rotate clockwise/counterclockwise, when viewed from above, and corroborated a weak hemispheric preference that both the filament and related sigmoid exhibit a forward (reverse) S-shape in the southern (northern) hemisphere. The one-to-one relationship between filament chirality and rotation direction may help understand the transformation of CMEs to interplanetary ejecta, and help predict statistically the latter's characteristics, such as the axial orientation, the chirality of helical fields, and the strength and duration of southward magnetic field. As illustrated in Figure~\ref{fig:figure4}, for both odd and even cycles, the rotation of eruptive active-region filaments makes their axes tilt toward the east-west direction. Such a statistical trend is already known for interplanetary magnetic clouds \citep{Crooker2000,Yurchyshyn2009}, yet its connection with filament rotation has not been noticed. On the other hand, Figure~\ref{fig:figure4} predicts that the axes of active-region filaments and consequently those of interplanetary ejecta are inclined toward the direction of the global solar field during the declining phase of each cycle, when the polar field polarity is consistent with that of follower sunspots---this prediction remains to be verified by observations.
%and make the latter's characteristics statistically predictable, such as the axial orientation, the chirality of helical fields, and the strength and duration of southward IMF. 

\acknowledgments
We acknowledge the SECCHI, AIA, GONG, XRT, and HMI consortia for providing the excellent observations. 
ZZ is supported by the Open Research Program of  CAS Key Laboratory of Geospace Environment.
RL and YW are supported by NSFC grants 41761134088, 41774150, 11925302, and 41774178.
LL is supported by the Open Project of CAS Key Laboratory of Geospace Environment, and NSFC grants 11803096. 
JC is supported by NSFC grants 41525015 and 41774186.
ZZ appreciates discussions and support with Dr. Q.H. Zhang. 
\software{SolarSoftWare \citep{Freeland_Handy_2012}}

 \end{CJK*}

\begin{thebibliography}{}
\expandafter\ifx\csname natexlab\endcsname\relax\def\natexlab#1{#1}\fi

\bibitem[{{Awasthi} {et~al.}(2018){Awasthi}, {Liu}, {Wang}, {Wang}, \&
  {Shen}}]{Awasthi2018}
{Awasthi}, A.~K., {Liu}, R., {Wang}, H., {Wang}, Y., \& {Shen}, C. 2018, \apj,
  857, 124

\bibitem[{{Bothmer} \& {Schwenn}(1994)}]{Bothmer&Schwenn1994}
{Bothmer}, V., \& {Schwenn}, R. 1994, \ssr, 70, 215

\bibitem[{{Canfield} {et~al.}(1999){Canfield}, {Hudson}, \&
  {McKenzie}}]{Canfield1999}
{Canfield}, R.~C., {Hudson}, H.~S., \& {McKenzie}, D.~E. 1999, \grl, 26, 627

\bibitem[{{Chen} {et~al.}(2014){Chen}, {Harra}, \& {Fang}}]{Chen_etal_2014}
{Chen}, P.~F., {Harra}, L.~K., \& {Fang}, C. 2014, \apj, 784, 50

\bibitem[{{Cheng} {et~al.}(2014){Cheng}, {Ding}, {Zhang}, {Srivastava}, {Guo},
  {Chen}, \& {Sun}}]{Cheng_etal_2014}
{Cheng}, X., {Ding}, M.~D., {Zhang}, J., {et~al.} 2014, \apjl, 789, L35

\bibitem[{{Crooker}(2000)}]{Crooker2000}
{Crooker}, N.~U. 2000, Journal of Atmospheric and Solar-Terrestrial Physics,
  62, 1071

\bibitem[{{Freeland} \& {Handy}(2012)}]{Freeland_Handy_2012}
{Freeland}, S.~L., \& {Handy}, B.~N. 2012, {SolarSoft: Programming and data
  analysis environment for solar physics}, , , ascl:1208.013

\bibitem[{{Gibson} {et~al.}(2002){Gibson}, {Fletcher}, {Del Zanna}, {Pike},
  {Mason}, {Mandrini}, {D{\'e}moulin}, {Gilbert}, {Burkepile}, {Holzer},
  {Alexander}, {Liu}, {Nitta}, {Qiu}, {Schmieder}, \& {Thompson}}]{Gibson2002}
{Gibson}, S.~E., {Fletcher}, L., {Del Zanna}, G., {et~al.} 2002, \apj, 574,
  1021

\bibitem[{{Gilbert} {et~al.}(2000){Gilbert}, {Holzer}, {Burkepile}, \&
  {Hundhausen}}]{Gilbert2000}
{Gilbert}, H.~R., {Holzer}, T.~E., {Burkepile}, J.~T., \& {Hundhausen}, A.~J.
  2000, \apj, 537, 503

\bibitem[{{Golub} {et~al.}(2007){Golub}, {Deluca}, {Austin}, {Bookbinder},
  {Caldwell}, {Cheimets}, {Cirtain}, {Cosmo}, {Reid}, {Sette}, {Weber},
  {Sakao}, {Kano}, {Shibasaki}, {Hara}, {Tsuneta}, {Kumagai}, {Tamura},
  {Shimojo}, {McCracken}, {Carpenter}, {Haight}, {Siler}, {Wright}, {Tucker},
  {Rutledge}, {Barbera}, {Peres}, \& {Varisco}}]{Golub_etal_2007}
{Golub}, L., {Deluca}, E., {Austin}, G., {et~al.} 2007, \solphys, 243, 63

\bibitem[{{Green} {et~al.}(2007){Green}, {Kliem}, {T{\"o}r{\"o}k}, {van
  Driel-Gesztelyi}, \& {Attrill}}]{Green_etal_2007}
{Green}, L.~M., {Kliem}, B., {T{\"o}r{\"o}k}, T., {van Driel-Gesztelyi}, L., \&
  {Attrill}, G.~D.~R. 2007, \solphys, 246, 365

\bibitem[{{Green} {et~al.}(2018){Green}, {T{\"o}r{\"o}k}, {Vr{\v{s}}nak},
  {Manchester}, \& {Veronig}}]{Green2018}
{Green}, L.~M., {T{\"o}r{\"o}k}, T., {Vr{\v{s}}nak}, B., {Manchester}, W., \&
  {Veronig}, A. 2018, \ssr, 214, 46

\bibitem[{{Guo} {et~al.}(2010){Guo}, {Schmieder}, {D{\'e}moulin}, {Wiegelmann},
  {Aulanier}, {T{\"o}r{\"o}k}, \& {Bommier}}]{Guo_etal_2010}
{Guo}, Y., {Schmieder}, B., {D{\'e}moulin}, P., {et~al.} 2010, \apj, 714, 343

\bibitem[{{Harvey} {et~al.}(2011){Harvey}, {Bolding}, {Clark}, {Hauth}, {Hill},
  {Kroll}, {Luis}, {Mills}, {Purdy}, {Henney}, {Holland}, \&
  {Winter}}]{Harvey_etal_2011}
{Harvey}, J.~W., {Bolding}, J., {Clark}, R., {et~al.} 2011, in AAS/Solar
  Physics Division Abstracts \#42, AAS/Solar Physics Division Meeting, 17.45

\bibitem[Isenberg \& Forbes(2007)]{Isenberg_Forbes_2007} Isenberg, P.~A., \& Forbes, T.~G.\ 2007, \apj, 670, 1453

\bibitem[{{Kaiser} {et~al.}(2008){Kaiser}, {Kucera}, {Davila}, {St.~Cyr},
  {Guhathakurta}, \& {Christian}}]{Kaiser2008}
{Kaiser}, M.~L., {Kucera}, T.~A., {Davila}, J.~M., {et~al.} 2008, \ssr, 136, 5

\bibitem[{{Kilpua} {et~al.}(2019){Kilpua}, {Lugaz}, {Mays}, \&
  {Temmer}}]{Kilpua2019}
{Kilpua}, E.~K.~J., {Lugaz}, N., {Mays}, M.~L., \& {Temmer}, M. 2019, Space
  Weather, 17, 498

\bibitem[{{Kliem} {et~al.}(2004){Kliem}, {Titov}, \&
  {T{\"o}r{\"o}k}}]{Kliem_etal_2004}
{Kliem}, B., {Titov}, V.~S., \& {T{\"o}r{\"o}k}, T. 2004, \aap, 413, L23

\bibitem[Kliem et al.(2012)]{Kliem_etal_2012} Kliem, B., T{\"o}r{\"o}k, T., \& Thompson, W.~T.\ 2012, \solphys, 281, 137


\bibitem[{{Kosugi} {et~al.}(2007){Kosugi}, {Matsuzaki}, {Sakao}, {Shimizu},
  {Sone}, {Tachikawa}, {Hashimoto}, {Minesugi}, {Ohnishi}, {Yamada}, {Tsuneta},
  {Hara}, {Ichimoto}, {Suematsu}, {Shimojo}, {Watanabe}, {Shimada}, {Davis},
  {Hill}, {Owens}, {Title}, {Culhane}, {Harra}, {Doschek}, \&
  {Golub}}]{Kosugi2007}
{Kosugi}, T., {Matsuzaki}, K., {Sakao}, T., {et~al.} 2007, \solphys, 243, 3

\bibitem[{{Kusano}(2005)}]{Kusano2005}
{Kusano}, K. 2005, \apj, 631, 1260

\bibitem[{{Leka} {et~al.}(2009){Leka}, {Barnes}, {Crouch}, {Metcalf}, {Gary},
  {Jing}, \& {Liu}}]{Leka_etal_2009}
{Leka}, K.~D., {Barnes}, G., {Crouch}, A.~D., {et~al.} 2009, \solphys, 260, 83

\bibitem[{{Lemen} {et~al.}(2012){Lemen}, {Title}, {Akin}, {Boerner}, {Chou},
  {Drake}, {Duncan}, {Edwards}, {Friedlaender}, {Heyman}, {Hurlburt}, {Katz},
  {Kushner}, {Levay}, {Lindgren}, {Mathur}, {McFeaters}, {Mitchell}, {Rehse},
  {Schrijver}, {Springer}, {Stern}, {Tarbell}, {Wuelser}, {Wolfson}, {Yanari},
  {Bookbinder}, {Cheimets}, {Caldwell}, {Deluca}, {Gates}, {Golub}, {Park},
  {Podgorski}, {Bush}, {Scherrer}, {Gummin}, {Smith}, {Auker}, {Jerram},
  {Pool}, {Soufli}, {Windt}, {Beardsley}, {Clapp}, {Lang}, \&
  {Waltham}}]{Lemen_etal_2012}
{Lemen}, J.~R., {Title}, A.~M., {Akin}, D.~J., {et~al.} 2012, \solphys, 275, 17

\bibitem[{{Liu} {et~al.}(2018){Liu}, {Chen}, \& {Wang}}]{Liu2018}
{Liu}, R., {Chen}, J., \& {Wang}, Y. 2018, Science China Physics, Mechanics,
  and Astronomy, 61, 69611

\bibitem[{{Liu} {et~al.}(2012){Liu}, {Kliem}, {T{\"o}r{\"o}k}, {Liu}, {Titov},
  {Lionello}, {Linker}, \& {Wang}}]{Liu_etal_2012}
{Liu}, R., {Kliem}, B., {T{\"o}r{\"o}k}, T., {et~al.} 2012, \apj, 756, 59

\bibitem[{{Liu} {et~al.}(2010){Liu}, {Liu}, {Wang}, {Deng}, \&
  {Wang}}]{Liu2010}
{Liu}, R., {Liu}, C., {Wang}, S., {Deng}, N., \& {Wang}, H. 2010, \apjl, 725,
  L84

\bibitem[{{Lynch} {et~al.}(2009){Lynch}, {Antiochos}, {Li}, {Luhmann}, \&
  {DeVore}}]{Lynch2009}
{Lynch}, B.~J., {Antiochos}, S.~K., {Li}, Y., {Luhmann}, J.~G., \& {DeVore},
  C.~R. 2009, \apj, 697, 1918

\bibitem[{{Martin}(1998)}]{Martin1998}
{Martin}, S.~F. 1998, \solphys, 182, 107

\bibitem[{{Martin} {et~al.}(1992){Martin}, {Marquette}, \&
  {Bilimoria}}]{Martin_etal_1992}
{Martin}, S.~F., {Marquette}, W.~H., \& {Bilimoria}, R. 1992, Astronomical
  Society of the Pacific Conference Series, Vol.~27, {The Solar Cycle Pattern
  in the Direction of the Magnetic Field along the Long Axes of Polar
  Filaments}, ed. K.~L. {Harvey}, 53


\bibitem[McCauley et al.(2015)]{McCauley_etal_2015} McCauley, P.~I., Su, Y.~N., Schanche, N., et al.\ 2015, \solphys, 290, 1703

\bibitem[{{Ouyang} {et~al.}(2017){Ouyang}, {Zhou}, {Chen}, \&
  {Fang}}]{Ouyang_etal_2017}
{Ouyang}, Y., {Zhou}, Y.~H., {Chen}, P.~F., \& {Fang}, C. 2017, \apj, 835, 94

\bibitem[{{Pesnell} {et~al.}(2012){Pesnell}, {Thompson}, \&
  {Chamberlin}}]{Pesnell_etal_2012}
{Pesnell}, W.~D., {Thompson}, B.~J., \& {Chamberlin}, P.~C. 2012, \solphys,
  275, 3

\bibitem[{Pevtsov(2002)}]{Pevtsov_2002}
Pevtsov, A.~A. 2002, solar physics, 207, 111

\bibitem[{{Pevtsov} {et~al.}(2003){Pevtsov}, {Balasubramaniam}, \&
  {Rogers}}]{Pevtsov_etal_2003}
{Pevtsov}, A.~A., {Balasubramaniam}, K.~S., \& {Rogers}, J.~W. 2003, \apj, 595,
  500

\bibitem[{{Pevtsov} {et~al.}(2014){Pevtsov}, {Berger}, {Nindos}, {Norton}, \&
  {van Driel-Gesztelyi}}]{Pevtsov2014}
{Pevtsov}, A.~A., {Berger}, M.~A., {Nindos}, A., {Norton}, A.~A., \& {van
  Driel-Gesztelyi}, L. 2014, \ssr, 186, 285

\bibitem[{{Romano} {et~al.}(2005){Romano}, {Contarino}, \&
  {Zuccarello}}]{Romano2005}
{Romano}, P., {Contarino}, L., \& {Zuccarello}, F. 2005, \aap, 433, 683

\bibitem[{Rust(1999)}]{Rust_1999}
Rust, D. 1999, GEOPHYSICAL MONOGRAPH-AMERICAN GEOPHYSICAL UNION, 111, 221

\bibitem[{{Rust} \& {Kumar}(1996)}]{Rust_Kumar_1996}
{Rust}, D.~M., \& {Kumar}, A. 1996, \apjl, 464, L199

\bibitem[{{Rust} \& {LaBonte}(2005)}]{Rust&Labonte2005}
{Rust}, D.~M., \& {LaBonte}, B.~J. 2005, \apjl, 622, L69

\bibitem[{{Rust} \& {Martin}(1994)}]{Rust_Martin_1994}
{Rust}, D.~M., \& {Martin}, S.~F. 1994, Astronomical Society of the Pacific
  Conference Series, Vol.~68, {A Correlation Between Sunspot Whirls and
  Filament Type}, ed. K.~S. {Balasubramaniam} \& G.~W. {Simon}, 337

\bibitem[{{Savcheva} {et~al.}(2014){Savcheva}, {McKillop}, {McCauley},
  {Hanson}, \& {DeLuca}}]{Savcheva_etal_2014}
{Savcheva}, A.~S., {McKillop}, S.~C., {McCauley}, P.~I., {Hanson}, E.~M., \&
  {DeLuca}, E.~E. 2014, \solphys, 289, 3297

\bibitem[{{Schou} {et~al.}(2012){Schou}, {Scherrer}, {Bush}, {Wachter},
  {Couvidat}, {Rabello-Soares}, {Bogart}, {Hoeksema}, {Liu}, {Duvall}, {Akin},
  {Allard}, {Miles}, {Rairden}, {Shine}, {Tarbell}, {Title}, {Wolfson},
  {Elmore}, {Norton}, \& {Tomczyk}}]{Schou_etal_2012}
{Schou}, J., {Scherrer}, P.~H., {Bush}, R.~I., {et~al.} 2012, \solphys, 275,
  229

\bibitem[{{Thompson} {et~al.}(2012){Thompson}, {Kliem}, \&
  {T{\"o}r{\"o}k}}]{Thompson2012}
{Thompson}, W.~T., {Kliem}, B., \& {T{\"o}r{\"o}k}, T. 2012, \solphys, 276, 241

\bibitem[{{Titov} \& {D{\'e}moulin}(1999)}]{Titov_Demoulin_1999}
{Titov}, V.~S., \& {D{\'e}moulin}, P. 1999, \aap, 351, 707

\bibitem[{{T{\"o}r{\"o}k} {et~al.}(2010){T{\"o}r{\"o}k}, {Berger}, \&
  {Kliem}}]{Torok_etal_2010}
{T{\"o}r{\"o}k}, T., {Berger}, M.~A., \& {Kliem}, B. 2010, \aap, 516, A49

\bibitem[{{Vourlidas} {et~al.}(2011){Vourlidas}, {Colaninno},
  {Nieves-Chinchilla}, \& {Stenborg}}]{Vourlidas2011}
{Vourlidas}, A., {Colaninno}, R., {Nieves-Chinchilla}, T., \& {Stenborg}, G.
  2011, \apjl, 733, L23

\bibitem[{{Yurchyshyn} {et~al.}(2009){Yurchyshyn}, {Abramenko}, \&
  {Tripathi}}]{Yurchyshyn2009}
{Yurchyshyn}, V., {Abramenko}, V., \& {Tripathi}, D. 2009, \apj, 705, 426

\bibitem[{{Yurchyshyn} {et~al.}(2001){Yurchyshyn}, {Wang}, {Goode}, \&
  {Deng}}]{Yurchyshyn2001}
{Yurchyshyn}, V.~B., {Wang}, H., {Goode}, P.~R., \& {Deng}, Y. 2001, \apj, 563,
  381

\bibitem[{{Zhou} {et~al.}(2017){Zhou}, {Zhang}, {Wang}, {Liu}, \&
  {Chintzoglou}}]{Zhou_etal_2017}
{Zhou}, Z., {Zhang}, J., {Wang}, Y., {Liu}, R., \& {Chintzoglou}, G. 2017,
  \apj, 851, 133

\bibitem[Zhou et al.(2019)]{Zhou_etal_2019} Zhou, Z., Cheng, X., Zhang, J., et al.\ 2019, \apjl, 877, L28



\end{thebibliography}
\end{document}